\documentclass[aps,prl,twocolumn,superscriptaddress,showpacs]{revtex4}
\usepackage{epsfig}
\begin{document}
\title{Density matrix renormalization group study of the charging of
a quantum dot strongly coupled to a single lead}
\author{Richard Berkovits}
\affiliation{The Minerva Center, Department of Physics,
    Bar-Ilan University, Ramat-Gan 52900, Israel}
\date{June 11, 2003, version 2.0}
\begin{abstract}
A new application of the density matrix renormalization group
(DMRG) method to a system composed of an interacting dot coupled
to a infinite one dimensional lead is presented. This method
enables one to study the influence of the coupling to an external
lead on the thermodynamical properties of the dot. It is shown
that this method reproduces known results for a non-interacting
dot coupled to a lead, i.e., that for strong coupling discrete
states remain in the dot. We show that these states are robust and
do not disappear once interactions in the dot are considered.
Moreover, due to these discrete states, Coulomb blockade affects
the charging of the dot even though its strongly coupled to a
lead.
\end{abstract}

\maketitle

\section{Introduction}

Recently there has been much interest in the influence of the
dot-lead couplings on the properties of a quantum dot. Once the
dot is strongly coupled to the lead a suppression of the Coulomb
blockade is predicted \cite{aleiner02}, arising from the fact that
the number of electrons in the dot is no longer a good quantum
number. Nevertheless, a weak remnant of the Coulomb blockade
physics persists even for strong coupling \cite{matveev95}.

Surprisingly, the details of the couplings to the lead play a very
important role in determining the properties of the dot.
K\"{o}nig, Gefen and Sch\"{o}n \cite{konig98} have shown
analytically that the tunnelling density of states of a
non-interacting two orbital dot strongly coupled to a single lead
contains a delta like peak. Such a feature indicates that a well
defined localized state remains in the dot although the dot is
strongly connected to the lead. Thus, if one measures the number
of electrons in the dot as function of the chemical potential or
gate voltage, one will observe sharp jumps once the chemical
potential crosses the localized state energy. Once two leads are
connected to a two orbital dot \cite{silva02}, a delta like peak
in the density of states of the dot exists if the lead-dot
coupling matrix elements are of equal signs, while no such feature
is predicted if one of the matrix elements is of opposite sign.
When the dot contains $N_{Dot}$ orbitals connected to a single
lead, $N_{Dot}-1$ delta like peaks appear in the tunnelling
density of states \cite{montambaux99}. The situation for
interacting dots has not been studied, although it was argued that
the delta like peaks will not be suppressed by interactions
\cite{silva02}.

Several numerical methods, especially exact diagonaliztion
\cite{berkovits98}, self consistent
Hartree-Fock\cite{levit99,walker99,bonci99} and density-functional
approaches\cite{hirose02,jiang03}, have proven themselves to be a
very useful tool in studying the properties of interacting quantum
dots\cite{alhassid00}. Unfortunately, none of these methods is
appropriate for the study of the strong lead-dot coupling regime.
Exact diagonalization methods are limited by the size of the
Hilbert space. Although one may consider a non-interacting lead,
due to the coupling with the interacting dot the whole system (dot
+ lead) must be treated via the full many particle states. This
will create an exponentially large Hilbert space even for a
moderate sized lead. Taking into account that the density of
states in the lead must be much larger than in the dot, its clear
that exact diagonalization is not a viable option. The treatment
of larger leads using self consistent Hartree-Fock is possible,
but at the price of using an uncontrollable approximation which
might lose important many body effects. Functional density methods
are not suitable for the treatment of steep potential gradients of
the type expected close to the dot-lead interface. Thus, a new
numerical approach is needed.

In this paper we shall use the density matrix renormalization
group method (DMRG) \cite{white93} in order to treat a single one
dimensional lead connected to a dot with $N_{Dot}$ orbitals. We
shall begin by describing the DMRG treatment of the lead-dot
system. We shall first review several features of the occupation
of the orbitals as function of the chemical potential for the
non-interacting case \cite{konig98,silva02,montambaux99}. Finally,
we shall establish that for an interacting dot all the $N_{Dot}-1$
states localized on the dot show full Coulomb blockade behavior in
the limit of strong coupling.

The paper is organized as follows: The second section contains the
definition of the dot-lead model. In the third section we describe
the details of the DMRG method developed for the calculation.
The orbital occupation as function of the chemical potential for
the non-interacting and interacting dots is studied in the forth
section. The results are discussed in the last section.

\section{Model}

The system which is discussed in this paper is a quantum dot
connected to a one dimensional lead. The electrons are treated as
spinless electrons. The quantum dot is represented by the
following Hamiltonian:
\begin{equation}
H_{QD}=\sum_{i=1}^{N_{Dot}} (\varepsilon_i-V_g) a_i^{\dag} a_i + U
\sum_{i>j}^N a_i^{\dag} a_i a_j^{\dag} a_j. \label{hqd}
\end{equation}
Here $a_i^{\dag}$ depicts the creation operator of an electron on
the i-th orbital in the dot, $\varepsilon_i$ corresponds to the
energy of the orbital, $N_{Dot}$ is the number of orbitals, and
$U=e^2/C$, where $C$ is the capacitance of the dot. The influence
of an external gate coupled to the dot is taken into account via
$V_g$. The one dimensional lead corresponds to:
\begin{equation}
H_{Lead}= -t \sum_{^j=1}^{\infty} c_j^{\dag} c_{j+1} +
h.c,\label{hlead}
\end{equation}
where $c_j^{\dag}$ is the creation operator of an electron on the
j-th site of the lead, and $t$ is the hopping matrix in the lead.
For an infinite lead with no coupling to the dot this will result
in a lead with a band width of $4t$. The tunnelling between the
dot and the lead is represented by:
\begin{equation}
H_{DL}= \sum_{i=1} V_i a_i^{\dag} c_{1} + h.c,\label{hdl}
\end{equation}
where the dot is assumed to be attached to the edge of the lead,
and the tunnelling amplitude between the i-th orbital in the dot
and the lead is given by $V_i$.

Since the lead is assumed to be infinite one can not assign a
fixed number of electrons in the system, as is customary in exact
diagonalization studies of interacting fermionic systems. The
number of electrons in our system is determined by the chemical
potential $\mu$. Thus, the Hamiltonian of the dot-lead system is
governed by:
\begin{equation}
H = H_{Dot} + H_{Lead} + H_{DL} - \mu \left( \sum_{i=1}^{N_{Dot}}
a_i^{\dag} a_i + \sum_{^j=1}^{\infty} c_j^{\dag} c_j
\right).\label{hamiltonian}
\end{equation}
Since we are interested in the number of electrons residing in the
dot at a given chemical potential $\mu$ (or at a particular gate
voltage $V_g$ for a given $\mu$) at zero temperature, we must
calculate the ground state eigenvector of $H$. In the next section
we shall discuss the numerical methods we use to calculate this
quantity.
\\

\section{Numerical Method}

As long as there are no interactions in the system (i.e., $U=0$ in
the dot) the main obstacle for obtaining the ground state of $H$
(Eq. \ref{hamiltonian}) is the infinite span of the lead. But
since we are really interested in the number of electrons in the
dot, one may expect that as long as level broadening of the
orbitals in the dot (roughly proportional to $\pi \nu |V_i|^2$ for
weak coupling, where $\nu$ is the local density of states in the
lead) is much larger than the level spacing within the lead, the
finite lead treatment of the system is accurate. Thus, replacing
the infinite lead by a finite one with $N$ sites makes sense as
long as $N \gg (4t/V_i)^2$ (where $\nu = 1/4t$). Taking a finite
lead coupled to the dot we can exactly diagonalize $H$ and obtain
the ground state for any value of $\mu$ of $V_g$.

\begin{figure}\centering
\epsfxsize8.5cm\epsfbox{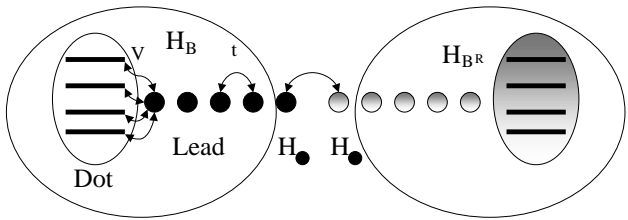} \vskip .5truecm \caption{The
structure of the superblock at the forth iteration. $H_B$
represents the dot and the lead. An additional site,
$H_{\bullet}$, is added to the lead. The whole left part ($H_{B
\bullet}$) is reflected ($H_{ \bullet B^R}$)and added to the right
of the block in order to form the superblock $H_{B \bullet \bullet
B^R}$.} \label{figrg}
\end{figure}

The situation is much more difficult when interactions in the dot
are taken into account. Exact diagonalization is out of questions,
since the size of the Hilbert space grows exponentially as
function of the number of sites in the lead. We will use the DMRG
method \cite{white93} to treat the system. This method is usually
used for one dimensional interacting system and the resulting
ground state is as accurate as the one obtained from exact
diagonalization. In order to incorporate the dot into the DMRG
method we propose the following procedure: First we create a block
composed of the dot and the first site of the lead,
\begin{equation}
H_{B} = H_{Dot} + H_{DL} - \mu \left( \sum_{i=1}^{N_{Dot}}
a_i^{\dag} a_i + c_1^{\dag} c_1 \right).\label{block}
\end{equation}
and exactly diagonalize it. Then we begin an iteration process. An
additional site (in the first iteration $j=2$) is added to the
lead, and the Hamiltonian is given by:
\begin{equation}
H_{B \bullet} = H_{B} + c_{j-1}^{\dag} c_j + h.c - \mu
c_j^{\dag} c_j.\label{blockb}
\end{equation}
Then a superblock Hamiltonian $H_{B \bullet \bullet B^R}$
is formed by joining $H_{B \bullet}$ to its mirror image
$H_{\bullet B^R}$ (see Fig. \ref{figrg}), resulting in
\begin{equation}
H_{B \bullet \bullet B^R} = H_{B \bullet} + c_j^{\dag} c_{j^R}
+ h.c + H_{\bullet B^R},\label{blockbbblock}
\end{equation}
where $j^R$ is the index of the mirror image lead. The ground
state $\Psi(B \bullet \bullet B^R)$, which is a function of the
superblock coordinates is calculated by the Lanczos method. Using
this ground state a density matrix,
\begin{equation}
\rho(B \bullet) = \sum_{\bullet B^R} \Psi(B \bullet \bullet B^R)
\Psi(B \bullet \bullet B^R),\label{dm}
\end{equation}
is formed (the trace is over the coordinates of the mirror
reflection part of the Hamiltonian $H_{B \bullet \bullet B^R}$).
The density matrix is diagonalized and only half of the states
with the largest eigenvalues are retained as a truncated basis. A
new block Hamiltonian $H_B$ is formed by projecting $H_{B
\bullet}$ onto the truncated basis. Also other relevant operators,
such as $a_i^{\dag}$, are rewritten in terms of the truncated
basis. The new $H_B$ now replaces the one given in Eq.
(\ref{block}) in a new iteration cycle. These iteration continue
until a satisfactory accuracy is obtained for the occupation of
the orbitals in the dot.
\\

\section{Results}

\subsection{Non interacting dot}

We shall begin by considering the behavior of a non-interacting
dot coupled to a one dimensional lead. We are interested in the
behavior of the orbital occupation in the dot as function of the
chemical potential $\mu$ or the gate voltage $V_g$. As discussed
in the previous section, we shall use numerical exact
diagonalization in order to obtain the single electron
eigenvectors $\phi_k$ and eigenvalues $\epsilon_k$ of the finite
lead coupled to a dot system, i.e. of $H$ as given in Eq.
(\ref{hamiltonian}). The dot's i-th orbital occupation, $n_i$, as
function of the chemical potential, is defined in the following
way:
\begin{equation}
n_i = \sum_k |\langle \phi_k|a_i^{\dag} a_i|\phi_k \rangle|^2
\theta(\mu-\epsilon_k).\label{ni}
\end{equation}
\begin{figure}\centering
\epsfxsize6.5cm\epsfbox{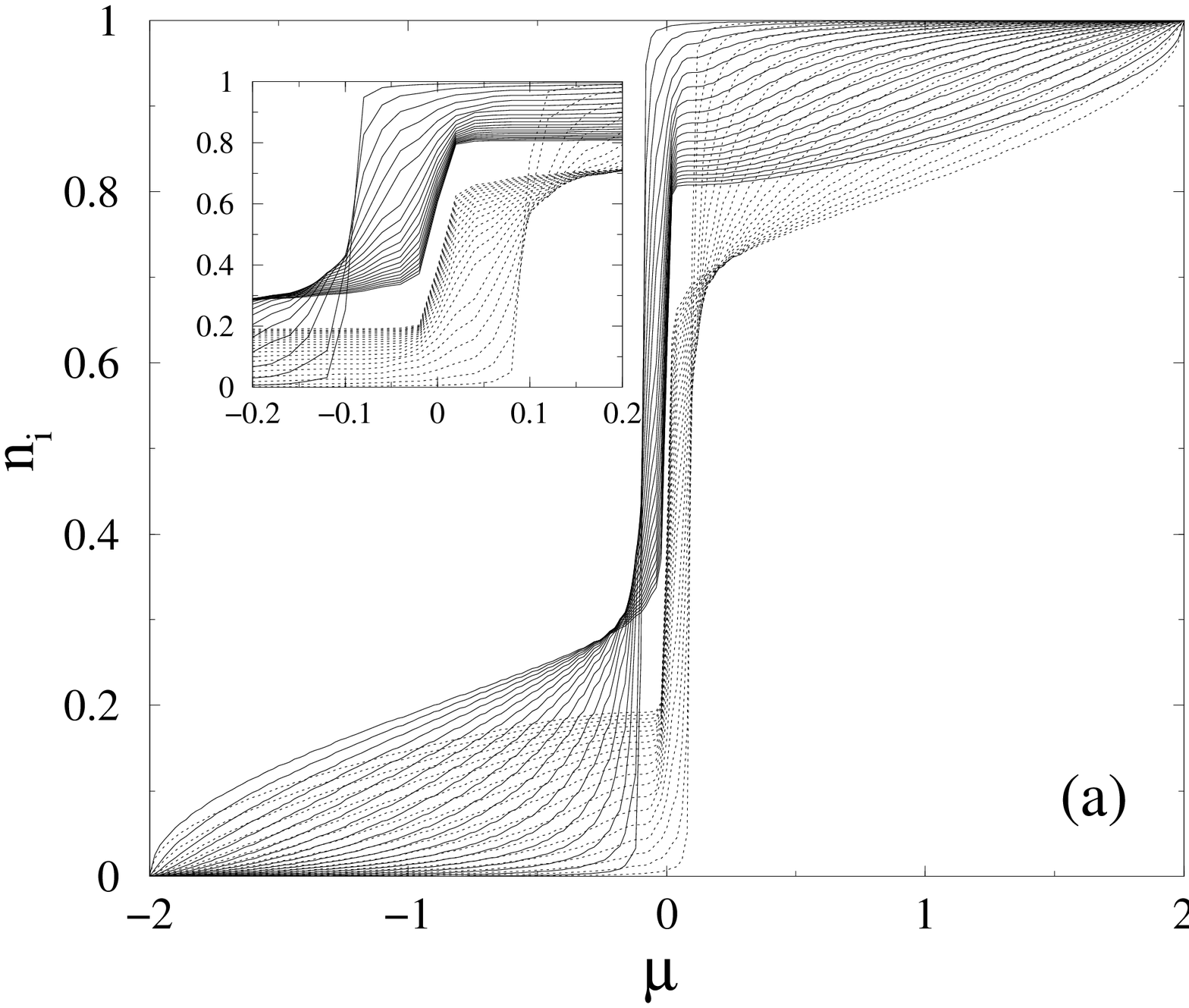}
\vskip -.5truecm
\epsfxsize6.5cm\epsfbox{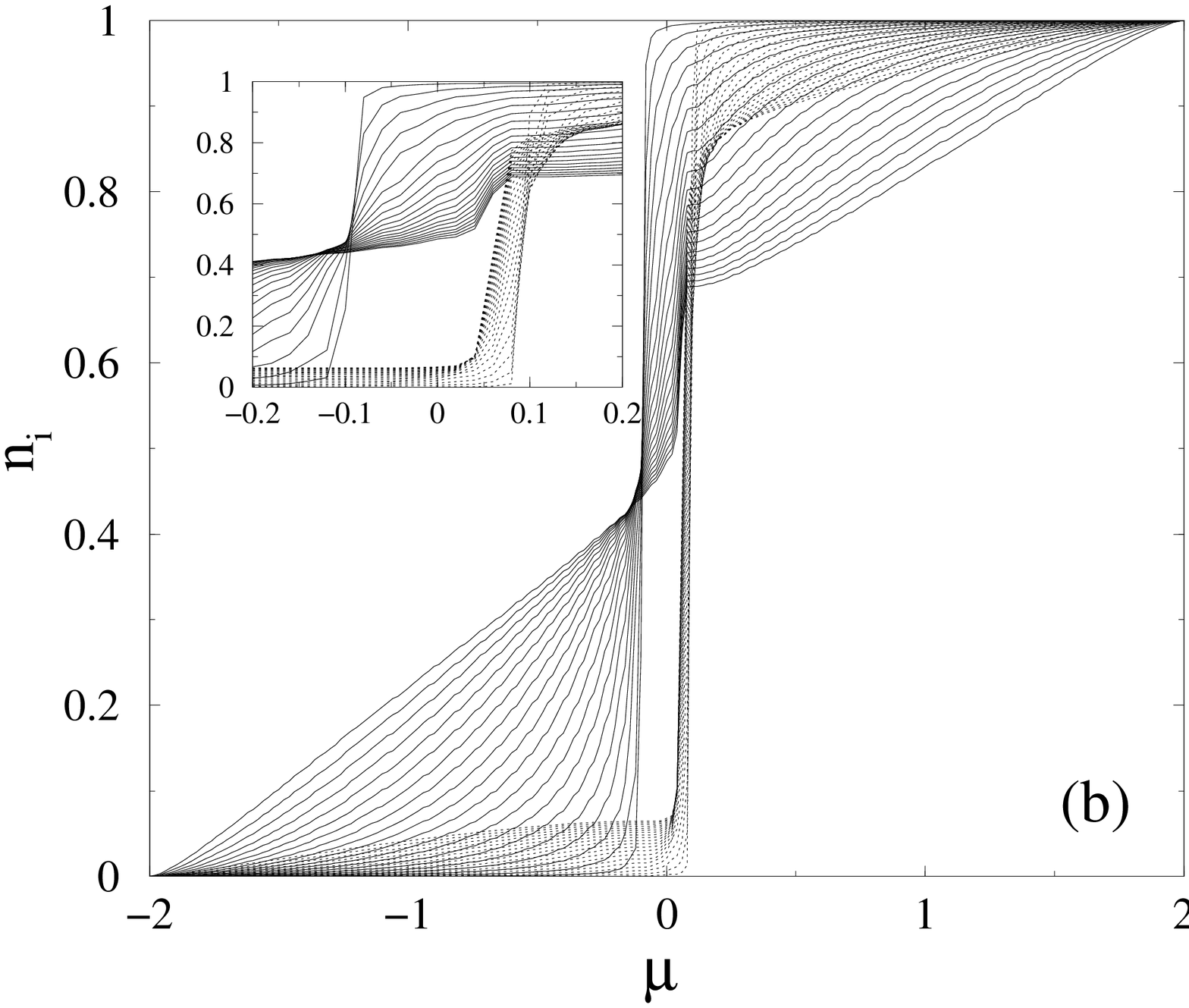} \caption{The orbital
occupation $n_1$ (full lines) and $n_2$ (dotted lines) of a two
orbital dot coupled to a one dimensional lead as function of $\mu$
for different values of the couplings $V_1$,$V_2$. In both plots
the series of curves correspond to different values of $V_1$ which
change between $V_1=0.05t$ (lowest curve at $\mu=-2t$) and $V_1=t$
(highest curve at $\mu=-2t$) in increments of $0.05t$. (a)
$V_2=V_1$; (b) $V_2=V_1/2$. Inset: An enlargement of the region
between $\mu=-0.2t$ and $\mu=0.2t$.} \label{fig2}
\end{figure}
In Fig. \ref{fig2}, typical results for $n_i(\mu)$ as function of
the coupling (a) $V_1=V_2=V$ and (b)$V_1=2V_2=V$ for a two orbital
dot are presented . As expected, for small values of $V$, $n_i$
shows a jump at $\varepsilon_i$, and this jump becomes less sharp
as $V$ increases. One might expect that for an increasing $V$, any
signature of the original orbitals in the dot will be washed out.
Nevertheless, quite surprising, that is not the case and as $V$
increases a {\it new} jump in both $n_1$ and $n_2$ appears for a
value of $\mu$ in between to original orbital energies
$\varepsilon_1$ and $\varepsilon_2$. The sum of the jumps in the
occupation of both orbital is equal to one for large $V$, i.e.,
although the dot is strongly coupled to the lead, a state which
contains a single electron remains localized in the dot. Such a
behavior was indeed predicted in Ref. \cite{konig98}, where the
tunnelling density of states for a non interacting two level
system coupled to a lead was calculated. They show that for strong
coupling the tunnelling density of states acquires a sharp peak at
an energy between the two levels, on the background of a very wide
peak. This behavior is reproduced when $\mu$ is kept constant but
the gate voltage coupled to the dot $V_g$ is changed.

\begin{figure}\centering
\epsfxsize6.5cm\epsfbox{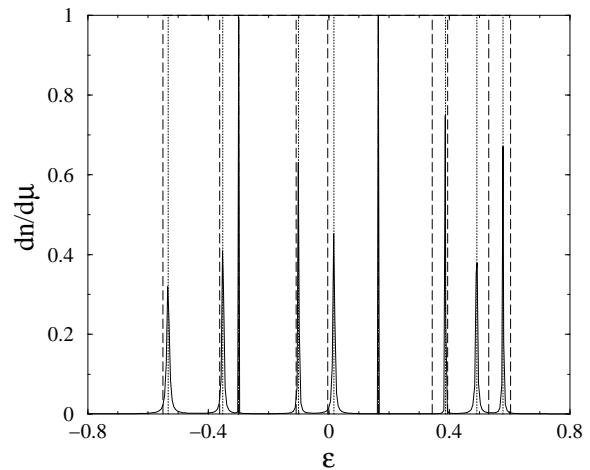} \caption{The derivative of
the total orbital occupation $\partial n/\partial \mu$ as function
of $\mu$ for a dot with $N_{Dot}=10$ described by a random matrix
Hamiltonian and random coupling $V_i$ between the dot and lead.
The dashed lines correspond to the case where $V_i=0$, the line
corresponds to an average value of the coupling $\langle V_i
\rangle = 0.4 t$, and the dotted line indicates the position of
the eigenvalues of Eq. (\ref{Nlevel}).} \label{fig4}
\end{figure}

As can be seen in Fig. \ref{fig2}, the location in $\mu$-space of
the localized state depends on the ratio of $V_1$ to $V_2$. This
location may be calculated using the following three state
Hamiltonian:
\begin{equation}
H_{3level}= \left(\matrix{\varepsilon_1 & 0 & V_1 \cr 0 &
\varepsilon_2 & V_2 \cr V_1 & V_2 & 0 \cr} \right), \label{3level}
\end{equation}
which takes into account the effect of the lead by a single site
that is strongly coupled to both orbitals. At the limit of large
coupling, $H_{3level}$ has two eigenvalues at $\pm
\sqrt{V_1^2+V_2^2}$, and (assuming $V_1>V_2$) a third at
\begin{equation}
\tilde{\varepsilon} = \varepsilon_2+\frac{\varepsilon_1 -
\varepsilon_2}{\left(\frac{V_1}{V_2}\right)^2+1}.\label{mid-level}
\end{equation}
The eigenvector of this state is composed exclusively from the two
dot orbitals, and the eigenvalue $\tilde{\varepsilon}$ corresponds
to the location of the jump in the orbital occupation seen in Fig.
\ref{fig2}. Thus, at the limit of strong coupling, the details of
the lead are not important in determining the energy of the
localized state on the dot, which can be adequately predicted by
coupling of all orbitals to a single external orbital
\cite{meir03}. This behavior is not limited to a two orbital dot
\cite{montambaux99}. For any number of orbitals $N_{Dot}$ in a dot
that are strongly coupled to the lead, the location of the
$N_{Dot}-1$ jumps in the orbital occupation may be estimated by
calculating the eigenvalues of an $N_{Dot}+1$ matrix:
\begin{equation}
\left(\matrix{\varepsilon_1 & 0 & 0 & . & V_1 \cr 0 &
\varepsilon_2 & 0 & . & V_2 \cr 0 & 0 & \varepsilon_2 & . & V_3
\cr . & . & . & . & .\cr V_1 & V_2 & V_3 & . & 0 \cr} \right).
\label{Nlevel}
\end{equation}
This is shown in Fig. \ref{fig4}, where the derivative of the
total orbital occupation as function of the chemical potential
$\partial n/\partial \mu$ (where $n = \sum_i n_i$) for a dot with
ten orbitals is compared with the eigenvalues of the matrix given
in Eq. (\ref{Nlevel}). An excellent correspondence for the
position of the $N_{Dot}-1$ peaks of $\partial n/\partial \mu$ and
the position of the intermediate $N_{Dot}-1$ eigenvalues of Eq.
(\ref{Nlevel}) is obtained.

\begin{figure}\centering
\epsfxsize6.5cm\epsfbox{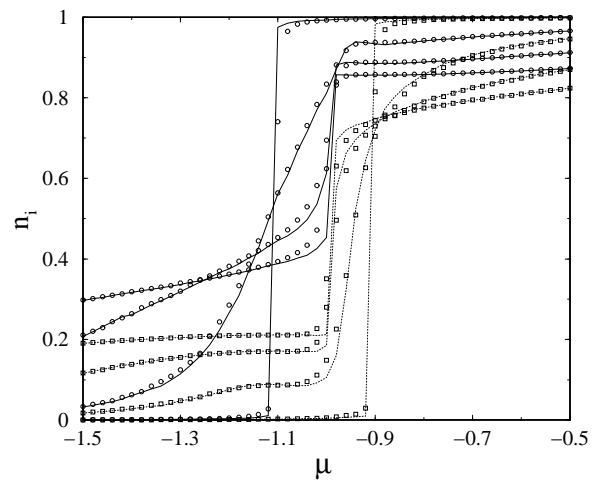} \caption{The orbital
occupation $n_1$ (full lines,circles) and $n_2$ (dotted
lines,squares) of a two orbital dot coupled to a one dimensional
lead as function of $\mu$ for different values of the couplings
$V_1=V_2=0.05t,0.3t,0.55t,0.8t$. The lines correspond to the DMRG
results while the symbols correspond to the exact diagonalization
results.} \label{fig5x}
\end{figure}

In the following section we shall discuss the role of interactions
in determining the orbital occupation of the dot. As previously
discussed, for the interacting case we must use the DMRG method.
It would be interesting nevertheless to compare the results of the
DMRG approach to exact diagonalization results. This, of course,
is only possible for non-interacting systems. An example of the
comparisons we performed are presented in Fig. \ref{fig5x}, where
the block size ($H_B$) was chosen to be $32$ states, resulting in
the maximum matrix size $H_{B \bullet \bullet B^R}$ of $4,096$.
The iterative process continued until the occupation of the levels
in the dot changed by no more than $10^{-3}$ for three consecutive
iterations. The correspondence between the exact diagonalization
results and the DMRG ones at the tails of the jumps is perfect. At
the immediate vicinity of the jump there is a slight difference
between the two methods. We shall return to the behavior of the
the DMRG in the immediate vicinity of the jump in the next
section.

\subsection{Interacting dot}

In this section we shall concentrate on the influence of
electron-electron interaction in the dot on the occupation of
levels in the dot coupled to a lead. For weak interactions we
expect the usual Coulomb blockade behavior, i.e., the position of
the jump in occupation of the j-th orbital (ordered according to
$\varepsilon_1<\varepsilon_2< \ldots < \varepsilon_{N_{Dot}}$)
should occur at $\mu=\varepsilon_j+(j-1)U$. Once the coupling with
the lead becomes strong enough, it is not clear what influence the
interactions will have on the existence of the localized states in
the dot discussed in the previous section.

\begin{figure}\centering
\epsfxsize6.5cm\epsfbox{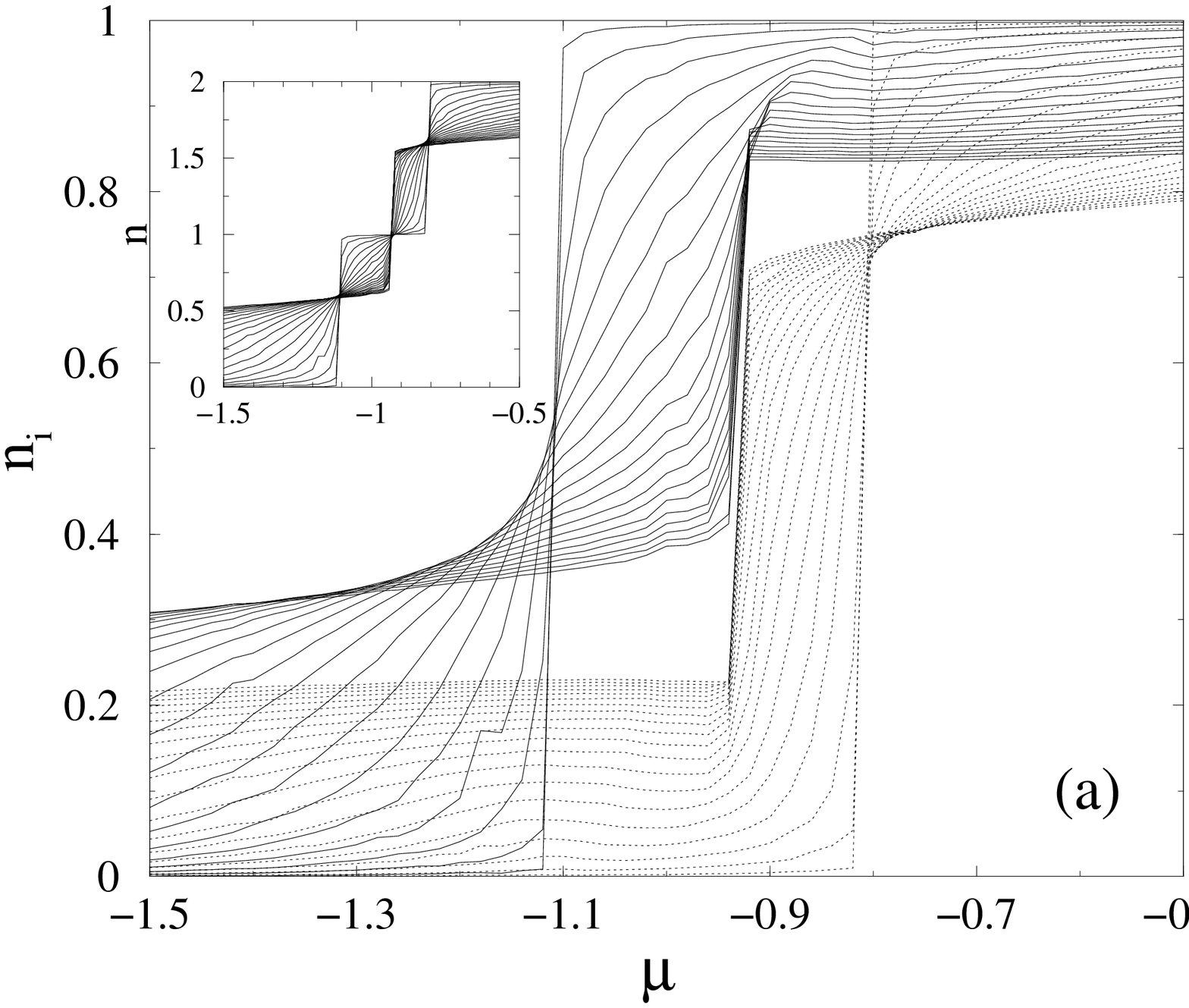}
\vskip -.5truecm
\epsfxsize6.5cm\epsfbox{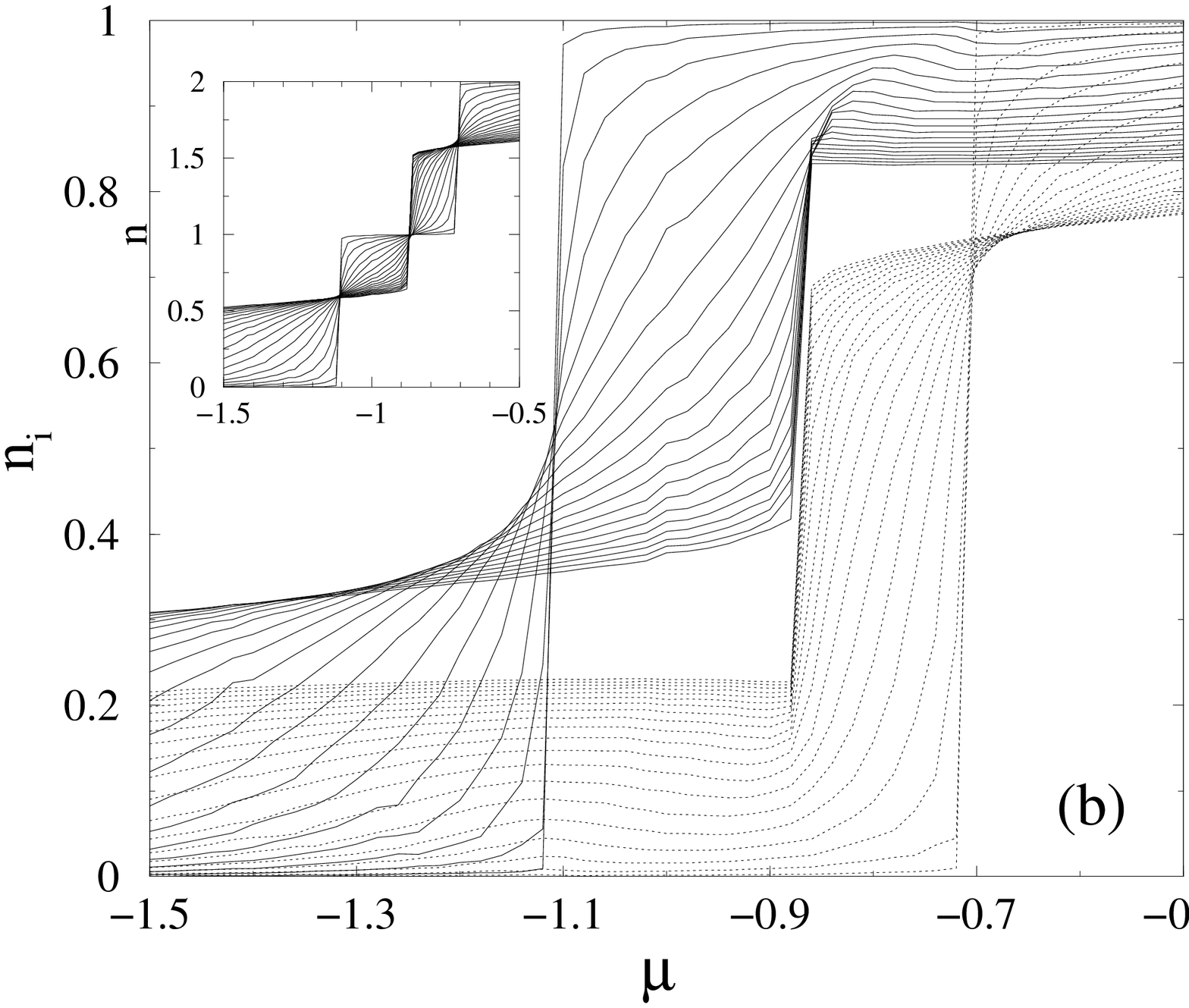}
\vskip -.5truecm
\epsfxsize6.5cm\epsfbox{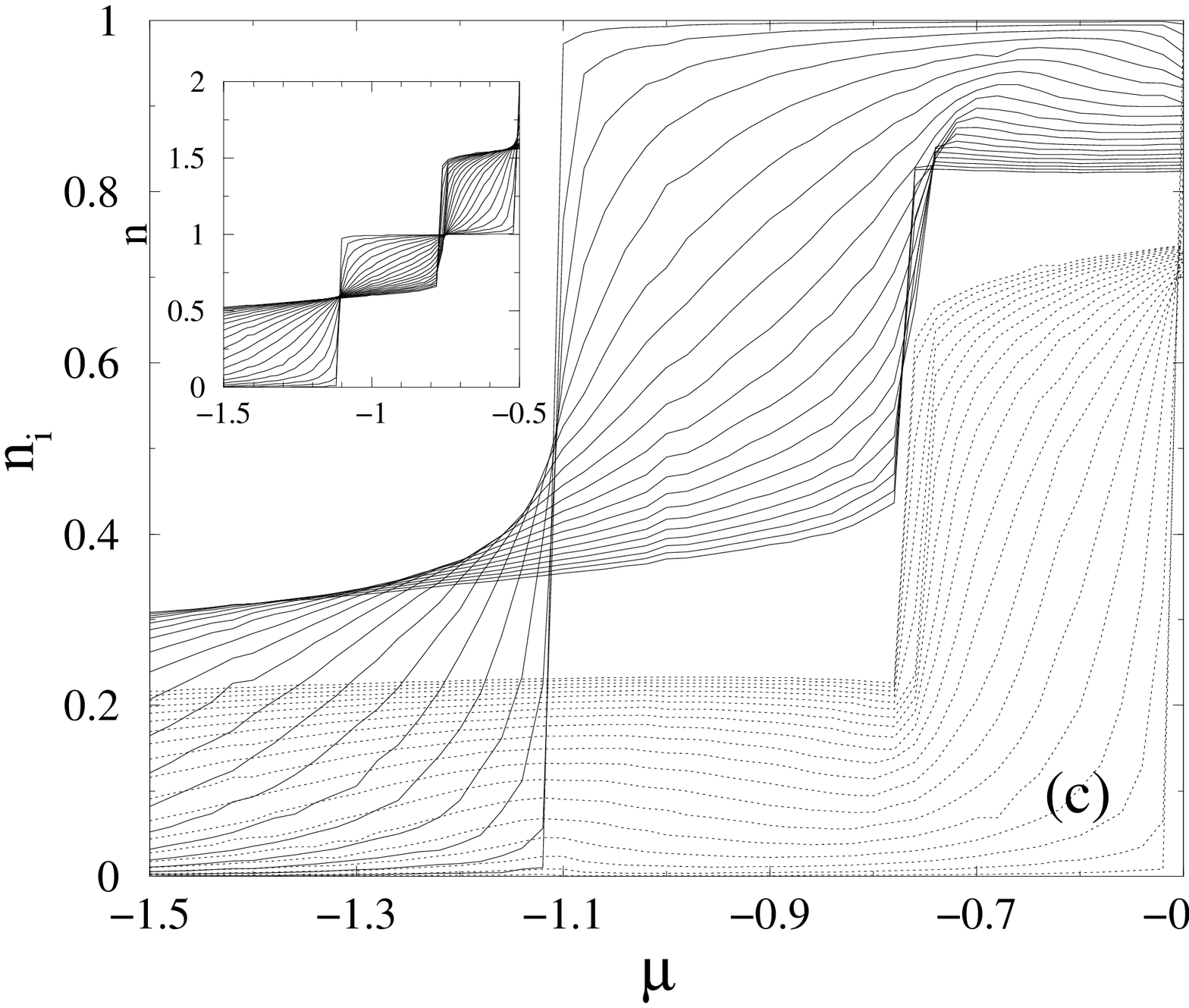} \caption{The orbital
occupation $n_1$ (full lines) and $n_2$ (dotted lines) of a two
orbital interacting dot coupled to a one dimensional lead as
function of $\mu$ for (a)$U=0.1t$; (b) $U=0.2t$; (c) $U=0.4t$. The
couplings $V_1=V_2=0.05t, 0.1t, \ldots t$. Inset: the total
orbital occupation $n=n_1+n_2$ as function of $\mu$.} \label{fig5}
\end{figure}

Let us begin by discussing the role of interactions on a two
orbital dot ($\varepsilon_1= -1.1t$, $\varepsilon_2= -0.9t$). The
orbital occupations for weak ($U=0.1t$), intermediate ($U=0.2t$)
and strong ($U=0.4t$) interactions are presented in Fig.
\ref{fig5}. The main feature of the non-interacting behavior,
i.e., that at the limit of strong coupling between dot and lead a
state which contains a single electron remains localized in the
dot (Fig. \ref{fig2}), persists in the interacting case. The
energy of the localized state corresponds to
$(\varepsilon_2+\varepsilon_1)/2+f(U)$, where $f(U)$ depends on
the change in occupation of both orbitals as the localized state
is filled. For $V_1=V_2$ one can see that $f(U)\sim U/2$, but we
shall analyze $f(U)$ in detail in the next paragraph. Another
interesting feature of interactions can be seen even for moderate
values of $V_i$. The orbital occupation of the second orbital
$n_2$, which is partially filled for low values of $\mu$
depopulate once a jump occurs in $n_1$. This is an obvious
response to the interaction between the two levels.

\begin{figure}\centering
\epsfxsize6.5cm\epsfbox{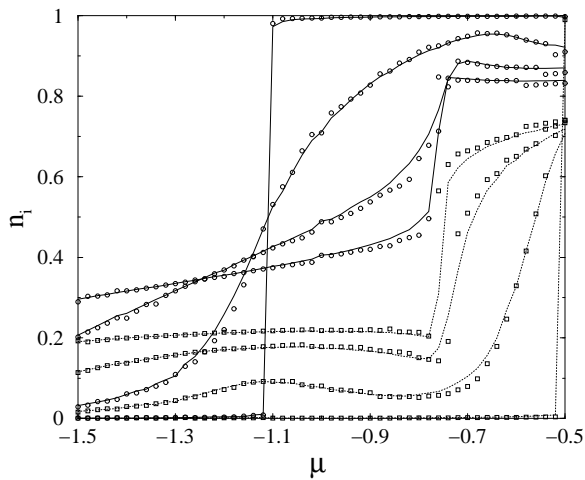} \caption{The orbital
occupation $n_1$ (full lines,circles) and $n_2$ (dotted
lines,squares) of a two orbital dot coupled to a one dimensional
lead as function of $\mu$ for different values of the couplings
$V_1=V_2=0.05t,0.3t,0.55t,0.8t$ at interaction strength $U=0.4t$
are compared. The lines correspond to the DMRG results with block
size of $32$ states while the symbols correspond to a block size
of $16$.} \label{fig10}
\end{figure}

Although it is not possible to directly compare the DMRG results
for the interacting case to an exact calculation, one can
nevertheless gain some insight into the reliability of the method
by checking its sensitivity to a change in the block size. In Fig.
\ref{fig10} the DMRG results for a two orbital dot in which a
block size of $16$ states is compared with a block size of $32$
states. Perfect correspondence between both sizes is seen at the
at the tails of the jumps, while small deviations are seen in the
immediate vicinity of the jump. This is very similar to the
situation for the non-interacting case (Fig. \ref{fig5x}). Thus,
the DMRG seems to be somewhat less accurate in the description of
the dot occupation in the immediate vicinity of a steep jump in
occupation.

\begin{figure}\centering
\epsfxsize6.5cm\epsfbox{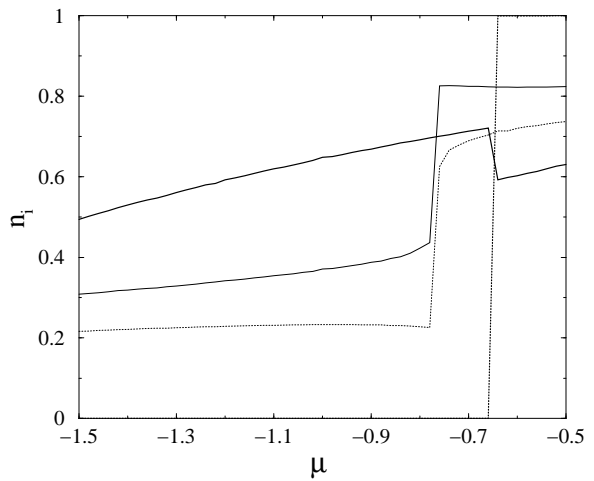} \caption{The orbital
occupation $n_1$ (full lines) and $n_2$ (dotted lines) of a two
orbital interacting dot ($U=0.2t$) coupled to a one dimensional
lead as function of $\mu$ for different couplings. In both cases
$V_1=1$, while the heavy lines correspond to $V_2=0$ and the
regular lines to $V_2=1$. } \label{fig5z}
\end{figure}

In order to clarify the the role played by the interaction and
coupling in determining $f(U)$, we begin by analyzing the case for
which $V_2=0$. i.e., the second orbital is disconnected from the
lead. Under such conditions the second orbital may acquire only
two values $n_2=0$ or $n_2=1$. Naively one expects that the second
orbital will be filled once $U n_1(\mu) < \mu-\varepsilon_2$. This
is indeed the case, but as can be seen in Fig. \ref{fig5z},
$n_1(\mu)$ is not a monotonous function of $\mu$, since once the
second orbital is populated, there is a reduction in the
occupation of the first orbital. Therefore, the condition for the
population of the second must be rewritten as $U n_1(\mu^+) <
\mu^+-\varepsilon_2$, where $\mu^+$ is the chemical potential just
above the jump. Of course, once the coupling $V_1$ is strong
enough to completely localize the remaining resonance, then
$n_1(\mu^+)=n_1(\mu^-)=1/2$ and the second orbital will populate
at $\mu=\varepsilon_2+U/2$. The situation becomes more complicated
once $V_2 \neq 0$. The additional charging energy due to the
population of the localized state is proportional to
$U(n_1(\mu^+)-n_1(\mu^-))(n_2(\mu^+)-n_2(\mu^-))$, which must be
smaller than $\mu-\tilde{\varepsilon}$ in the limit of strong
coupling.

\begin{figure}\centering
\epsfxsize6.5cm\epsfbox{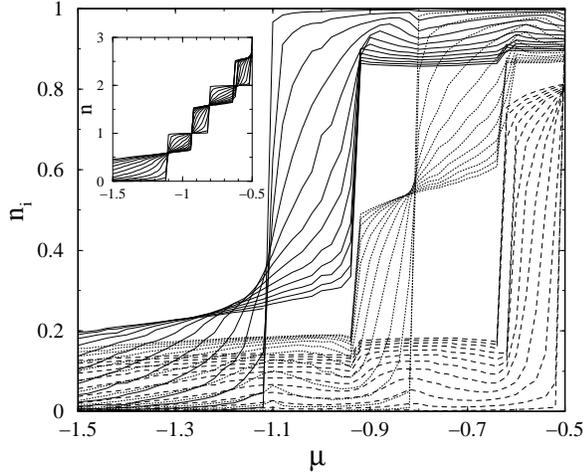} \caption{The orbital
occupation $n_1$ (full lines), $n_2$ (dotted lines), and $n_3$
(dashed lines) of a three orbital interacting dot ($U=0.2t$)
coupled to a one dimensional lead as function of $\mu$. The
couplings $V_1=V_2=V_3=0.05t, 0.1t, \ldots 0.6t$.Inset: the total
orbital occupation $n=n_1+n_2+n_3$ as function of $\mu$. }
\label{fig6}
\end{figure}

\begin{figure}\centering
\epsfxsize6.5cm\epsfbox{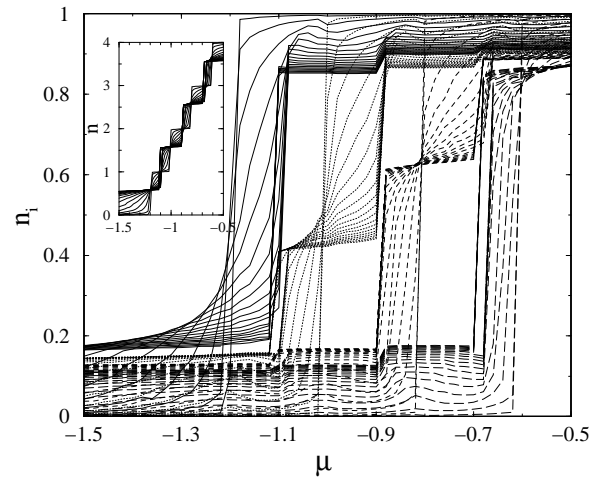} \caption{The orbital
occupation $n_1$ (full lines) and $n_2$ (dotted lines), $n_3$
(dashed lines),$n_4$ (long-dashed lines) of a four orbital
interacting dot ($U=0.1t$) coupled to a one dimensional lead as
function of $\mu$ for couplings $V_1=V_2=V_3=V_4=0.05t, 0.1t,
\ldots 0.6t$. Inset: the total orbital occupation
$n=n_1+n_2+n_3+n_4$ as function of $\mu$. } \label{fig7}
\end{figure}

As might be expected, the situation is similar for dots with a
higher number of orbitals. A three orbital dot is depicted in Fig.
\ref{fig6}. The orbitals have energies $\varepsilon_1=-1.1$,
$\varepsilon_2=-1$,$\varepsilon_3=-0.9$, and the interaction
$U=0.2$. In Fig. \ref{fig7} a four orbital dot, with energies
$\varepsilon_1=-1.2$, $\varepsilon_2=-1.1$,$\varepsilon_3=-1$,
$\varepsilon_4=-0.9$, and interaction $U=0.1$ is shown. In both
cases there is a transition from $N_{Dot}$ jumps in the occupation
of the dot for weak couplings to $N_{Dot}-1$ jumps for strong
couplings. The jumps in the strong coupling regime are separated
by $U+\Delta$ ($\Delta$ is the orbital spacing) as in the regular
Coulomb blockade case \cite{alhassid00}. The suppression of the
the occupation of higher orbitals as lower ones become occupied is
also evident. Thus, the Coulomb blockade behavior between the
resonances which remain localized on the dot continues although
the dot is well connected to the external lead.

\section{Discussion}

As we have seen in the previous section, the DMRG method enables
us to calculate the charging of a dot as function of the chemical
potential for the whole span of coupling strength. Contrary to
intuition, strong coupling to a single one dimensional lead does
not remove all discrete features from the dot, as was discussed in
Refs. \cite{konig98,silva02,montambaux99}. This conclusion remains
valid also for a dot in which interactions between electrons are
taken into account. These interactions manifest themselves as
Coulomb blockade steps in the filling of the dot.

This paper concentrated on developing the DMRG numerical method
and examining the role of interaction at the strong coupling
limit. Several important questions remain open. The crossover from
the weak coupling to the strong coupling regime should be
quantified, and its dependence on the level spacing in the dot and
the interaction studied. This is especially important since Ref.
\cite{aleiner02}, which essentially treats the intermediate values
of coupling, sees no pronounced discrete features in the dot.
Thus, one may speculate that an intermediate regime of coupling,
for which discrete features in the dot are suppressed, exists
between the weak and strong coupling regimes in which these
features are pronounced. Hints of this intermediate regime might
be seen in Figs. (\ref{fig6},\ref{fig7}) for $V_1 \sim 0.2$, but
this warrants further study. The influence of the dimensionality
of the lead, the number of channels in the lead and the number of
leads should be clarified. The DMRG method presented in this paper
could, in principal, be expanded to deal with several
one-dimensional leads, and probably also to quasi-one dimensional
leads. Thus, by expanding the methods introduced here, it may be
possible to answer open questions regarding the dot-lead coupling.

Very useful discussions with B. L. Altshuler, Y. Gefen, J.
Kantelhardt, Y. Meir and F. von Oppen are gratefully acknowledged,
as well as support from the Israel Academy of Science.

\end{document}